\documentclass[aps,prd,twocolumn,showpacs,eqsecnum,amsmath,nofootinbib,floatfix,superscriptaddress]{revtex4}
\usepackage[T1]{fontenc}
\usepackage[utf8]{inputenc}
\usepackage{lmodern}
\usepackage[english]{babel}
\usepackage{xcolor}
\usepackage{graphicx,amssymb,amsthm,amsmath,mathrsfs}
\usepackage{hyperref}
\def \d {{\rm d}}
\def \e {e}

\newtheorem{conjecture}{Conjecture}

\begin{document}

\title{Well-posed non-vacuum solutions in Robinson--Trautman geometry}
\author{T. Tahamtan}
\email[]{tahamtan@utf.mff.cuni.cz}
\affiliation{Institute of Theoretical Physics, Faculty of Mathematics and Physics, Charles University, Prague, V~Hole\v{s}ovi\v{c}k\'ach 2, 180~00 Prague 8, Czech Republic}

\author{D. \surname{Flores-Alfonso}}
\email[]{danflores@unap.cl}
\affiliation{Instituto de Ciencias Exactas y Naturales, Universidad Arturo Prat, Avenida Playa Brava 3256, 1111346, Iquique, Chile}
\affiliation{Facultad de Ciencias, Universidad Arturo Prat, Avenida Arturo Prat Chac\'on 2120, 1110939, Iquique, Chile}

\author{O. Sv\'{\i}tek}
\email[]{ota@matfyz.cz}
\affiliation{Institute of Theoretical Physics, Faculty of Mathematics and Physics, Charles University, Prague, V~Hole\v{s}ovi\v{c}k\'ach 2, 180~00 Prague 8, Czech Republic}

\begin{abstract}
We study nonlinear matter models compatible with radiative Robinson--Trautman spacetimes and analyze their stability and well-posedness. The results lead us to formulate a conjecture relating the (in)stability and well/ill-posedness to the character of singularity appearing in the solutions. We consider two types of nonlinear electrodynamics models, namely we provide a radiative ModMax solution and extend recent results for the RegMax model by considering the magnetically charged case. In both cases, we investigate linear perturbations around stationary spherically symmetric solutions to determine the stability and principal symbol of the system to argue about well-posedness of these geometries. Additionally, we consider a nonlinear sigma model as a source for Robinson--Trautman geometry. This leads to stationary solutions with toroidal (as opposed to spherical) topology thus demanding modification of the analysis.
\end{abstract}

\date{\today}

\maketitle

\section{Introduction}

Nonlinear Electrodynamics (NE) is a class of theories which generalize Maxwell's equations by departing from the superposition principle. Examples of such theories are known to emerge from quantum field theory, meaning that the electromagnetic properties of the quantum vacuum are effectively described by nonlinear polarization and magnetization relations. This perspective, of treating empty space as a medium, has led to phenomena such as the scattering of light by light and the deflection of light in electric fields~\cite{Euler:1935zz,Weisskopf:1936hya}.

In their seminal work, Born and Infeld noticed that the infinite self-energy of a point charge can be rendered non-singular if one allows for nonlinear constitutive relations~\cite{BornInfeld}. This regularity persists when self-gravity is taken into account. Take, for instance, the Reissner--Nordstr\"{o}m black hole which has a field singularity where the curvature of the background becomes infinite. In that case, the electromagnetic divergence lies behind the spacetime's event horizon. If instead, the Born--Infeld constitutive relation is used then there is no such field divergence, nonetheless, the spacetime singularity remains~\cite{Salazar:1984}. 


In a static and spherically symmetric configuration, however, other NE models can produce so called  regular black holes where the spacetime singularity is removed due to the properties of the source, the most famous being~\cite{Ayon-Beato:1998hmi}. Nevertheless, while plenty of gravitating NE systems resolve singularities in the electromagnetic sector, only a limited set of models resolve the curvature singularity. It is thus natural to question if regular black holes are an outcome of high symmetry of spherical solutions. The first results that suggest this might be the case were found in Ref.~\cite{Tahamtan-NE:2016}. There, a generating technique was established that constructs Robinson--Trautman (RT) geometries out of spherically symmetric seeds. The method was explicitly shown to generate a variety of RT-NE configurations yet was unsuccessful for regular black holes. What is more, it was recently shown that radiative configurations cannot be sourced by arbitrary NE theories~\cite{Tahamtan-PRD:2021}. An important caveat to these results, however, is that the RT class includes static spherically symmetric spacetimes but excludes the metrics of rotating black holes, by definition. Nonetheless, no rotating regular black holes have been reported as of yet and there are also no significant results for generalization of the RT family which includes rotation.

%

Let us recall that the geometries which belong to the Robinson--Trautman family of spacetimes are all those which admit a geodesic null congruence with vanishing twist and shear but with nontrivial expansion. Generic members of this family contain exact gravitational radiation. These spacetimes generalize static black holes in such a way that they have found application in astrophysics, particularly, in the study of binary black hole mergers~\cite{Rezzolla:2010df}. The ringdown phase of these systems can be analyzed with black hole perturbation theory, however, the sudden deceleration ("antikick") of the recoiling black hole can be understood employing RT metrics. Since binary systems of compact objects are an important source of gravitational waves then radiating RT configurations represent an interesting spacetime geometry; at the very least the metrics represent excellent toy models~\cite{deOliveira:2011pk,Jaramillo:2011re}. Additionally, when coupled to Maxwellian fields the configurations radiate, in general, both gravitational and electromagnetic waves.

In this paper, we study new radiative Robinson--Trautman configurations sourced by nonlinear electrodynamics. Finding such dynamical spacetimes is not an easy task as they solve Einstein's equations exactly with little to no symmetries. The inherent difficulties involved are partly reflected in the scarce amount of examples found in the literature for NE matter sources. There is no doubt that in general relativity the Robinson--Trautman subclass of vacuum solutions represents a fundamental family of metrics containing a wide variety of important configurations. In the past several decades, RT geometries have been studied within several different contexts which include black hole formation, gravitational radiation, generalized hairy black holes and holographic applications ~\cite{Chrusciel:1991vxx,Chrusciel:1992tj,Bicak:1995vc,Gueven:1996zm,Bakas:2014kfa,Tahamtan-PRD:2015,Tahamtan-PRD:2016,Marcello-sigma}. On the other hand, nonlinear electrodynamics presents its own challenges when self-gravitating systems are under consideration. Indeed, in most cases high symmetry plays a key role in solving the equations of motion; to see this we suggest the following Refs.~\cite{Gibbons:2001sx,Burinskii:2002pz,Elizalde:2003ku,Aiello:2004rz,Tahamtan-fR-MAX:2012,Alvarez:2014pra,Canate:2020btq,Flores-Alfonso:2020nnd,Alvarez:2022upr, Tahamtan-wormhole:2018,Tahamtan2023}. Those works study nonlinearly charged universes, black holes and wormholes with various kinds of asymptotic behavior. Some of them are framed within Einstein's theory of gravity, while others go beyond and consider more general types of dynamics. In particular, general relativity in spacetime dimension other than four has been the arena for both RT metrics and NE sources, studied separately in Refs.~\cite{Svitek2009,Hassaine:2007py,Miskovic:2010ey,Cardenas:2014kaa}. Broadly speaking, the picture painted by these investigations is that four dimensions is special for both subject matters as richer structures are available. However, going beyond linear constitutive relations in higher dimensions enhances RT configurations notably, even to the point of admitting radiative spacetimes, as showcased in Ref.~\cite{Kokoska:2021lrn}. The defining property of the NE models employed in that work was conformal symmetry, a point to which we return further below.

The results described in the following pages lead us to formulate the following conjecture:
\begin{conjecture}\label{conjecture}
The RT solutions are stable and well-posed around static black hole solutions which have spacelike singularities, while they are unstable and ill-posed in the case of timelike singularity.
\end{conjecture}

\section{The Matter Sources}

In this manuscript, we opt to work in a four-dimensional scenario as it provides us with the most general dynamical geometries within the RT class and the richest theories within the NE class. The latter of these facts is related to the number of independent Lorentz invariants which can be constructed from the field strength, $\textbf{F}$. To further elaborate, recall that in four dimensions this number is two and the standard relativistic invariants used are
\begin{align}\label{invariants}
 \mathcal{F}=F_{\mu\nu}F^{\mu\nu}, \qquad \text{and} \qquad \mathcal{G}=F_{\mu\nu}{}^{*}F^{\mu\nu},
\end{align}
where ${}^*$ is indicative of the Hodge linear map. Notice that the pseudoscalar $\mathcal{G}$ is particular to four dimensions, i.e., it has no analogue in other dimensions. Maxwell theory, for instance, can be formulated by a Lagrangian that is independent of $\mathcal{G}$, thus it can be straightforwardly generalized to other dimensions. This Lagrangian formulation is convenient as it is manifestly Lorentz invariant. In accordance with this, various NE theories can be formulated by employing functions of the standard invariants  $\mathbb{L}(\mathcal{F},\mathcal{G})$. Among these Lagrangians we find the important subfamily described by functions of $\mathcal{F}$ only, say $\mathbb{L}(\mathcal{F})$. Relevant examples of such theories include the conformal power-Maxwell models referenced earlier, the only conformal electrodynamics within the subclass. Since they are indexed by the spacetime dimension then there is only one possible conformal power-Maxwell model for any given dimension; which in four dimensions is the linear Maxwell case. Notwithstanding, beyond the $\mathbb{L}(\mathcal{F})$ class there are infinitely many conformal Lagrangian theories; those of the form $\mathbb{L}(\mathcal{F},\mathcal{G})=\mathcal{F}f(\mathcal{G}/\mathcal{F})$. This showcases how much richer $\mathbb{L}(\mathcal{F},\mathcal{G})$ theories are when compared to the $\mathbb{L}(\mathcal{F})$ subclass. 


In this work, we assume that the nonlinear electrodynamics models are described via the Lagrangian which is a function of the invariants \eqref{invariants}. Hence, we consider the action~\cite{Plebanski:1970}
\begin{equation}\label{action}
	S=\frac{1}{2}\int d^{4}x \sqrt{-g}\,\left[\mathbb{R}+\mathbb{L}(\mathcal{F},\mathcal{G})\right]\, ,
\end{equation}
where $\mathbb{R}$ is the Ricci scalar for the metric $g_{\mu \nu}$ and $\mathbb{L}$ is the Lagrangian describing a NE model\footnote{Metric signature is $(-+++)$ and geometrised units are used in which $c=\hbar=8 \pi G=1$ and partial derivative is denoted by a comma.}. By varying the functional of Eq. \eqref{action} with respect to the metric, we get Einstein equations
\begin{equation}\label{field equations}
	G^{\mu}{}_{\nu}=T^{\mu}{}_{\nu}\ ,
\end{equation}
with an energy momentum tensor
\begin{equation}\label{energy-momentum-Maxwell}
T^{\mu}{}_{\nu}=\frac{1}{2}\{\delta ^{\mu}{}_{\nu}\,(\mathbb{L}-\mathcal{G}\,\mathbb{L}_{\mathcal{G}})-4\,(F_{\nu \lambda}F^{\mu \lambda})\mathbb{L}_\mathcal{F}\}.
\end{equation}
Here we have used the abbreviations $\mathbb{L}_\mathcal{F}=\frac{\partial\,\mathbb{L}}{\partial \mathcal{F}}$,  $\mathbb{L}_\mathcal{G}=\frac{\partial\,\mathbb{L}}{\partial \mathcal{G}}$, $\mathbb{L}_{\mathcal{FF}}=\frac{\partial^{2} \mathbb{L}}{\partial^2 \mathcal{F}}$, etc.
In addition, the matter equations read
\begin{eqnarray}
F_{[\mu \nu,\lambda]}=0, \label{NE-MAX-TEN-2}\\
\left(\sqrt{-g}\,\mathbb{L}_{\mathcal{F}}\,F^{\mu\nu}+\sqrt{-g}\,\mathbb{L}_{\mathcal{G}}\,{}^{*}F^{\mu \nu}\right)_{,\mu}=0. \label{NE-MAX-TEN-1}
\end{eqnarray}

A significant result, that was recently established in Ref.~\cite{Tahamtan-PRD:2021}, is that there is only one  $\mathbb{L}(\mathcal{F})$ theory for which electrically charged radiative spacetimes exist with Robinson--Trautman geometry,
namely,
\begin{equation}\label{NewLag}
	\mathbb{L}=-8\,\alpha^4\,\left(1-3\ln(1-s)+\frac{s^3+3s^2-4s-2}{2(1-s)}\right)
\end{equation}
where $s={\left(\frac{-\mathcal{F}}{2\,\alpha ^4}\right)^\frac{1}{4}}$.
Moreover, unlike for Maxwell matter, for this particular NE source electrically charged RT geometries represent well-posed problems. This theory was later dubbed \emph{RegMax} and a wide variety of spacetime geometries sourced by it were found in Ref.~\cite{Tahamtan2023}. What is more, it was shown to provide slowly rotating black holes that can be found in a form naturally generalizing the slowly rotating Kerr--Newman configuration~\cite{Tahamtan-PRD:2022}. 

The RegMax electrodynamics was formulated with electrically charged spacetimes in mind, however, magnetically charged solutions may also be of interest. Since the theory belongs to the $\mathbb{L}(\mathcal{F})$ subclass it must lack electric/magnetic duality rotation invariance. This is because imposing the symmetry on the subfamily results in linear constitutive relations. Thus, electric and magnetic configurations are, in general, separate entities for $\mathbb{L}(\mathcal{F})$ theories. Magnetically charged spherically symmetric black holes have already been compared to their electrical counterpart, see Ref.~\cite{Tahamtan2023}. In this paper, we construct radiating magnetic configurations in RT geometry in order to examine their stability and well-posedness.

Now, it should be mentioned that finding exact solutions of Einstein's equations, albeit time-dependent solutions (e.g. the RT spacetimes), is not itself a guarantee that they are physically relevant. Physical properties of RT vacuum solutions based on several stability theorems have been known for decades, see for example Ref.~\cite{Dray:1982ex} for early work. Moreover, vacuum RT configurations asymptotically decay into the Schwarzschild black hole; a result that was proved analytically in Refs.~\cite{Chrusciel:1992tj,Chrusciel:1991vxx}. Yet when matter sources are taken into account this is not always the case, with Maxwell fields a key example. Lun and Chow studied linear perturbations around the Reissner--Nordstr\"{o}m solution within the RT-Maxwell system and showed they generally diverge exponentially in Ref.~\cite{Lun:1994up}. Later on, it was shown in Ref.~\cite{Kozameh_2008} that the system not only has an unstable branch but also that the equations do not constitute a well-posed initial value problem. This motivated the exploration of more general electrodynamics, as it would shine a light on whether all types of electromagnetic radiative solutions within the RT class suffer from this problem~\cite{Tahamtan-PRD:2021}. The result was the RegMax electrodynamics for which electric radiative RT spacetimes are stable and well posed.

As explained above, the only way to further explore radiative systems with Robinson--Trautman geometry is to go beyond the $\mathbb{L}(\mathcal{F})$ subclass, thus, we consider Lagrangians of a more general type. The only $\mathbb{L}(\mathcal{F},\mathcal{G})$ model to have already been studied in the literature, within this context, is that of Born--Infeld, for which the absence of radiative configurations was established in Ref.~\cite{Tahamtan-PRD:2021}. The Born--Infeld electrodynamics shares a few similarities with Maxwell's, for instance, birefringence is absent, the equations of motion are duality rotation invariant and the Lagrangian is Legendre self-dual~\cite{Gibbons:2000xe}. Notwithstanding, the defining feature of that electrodynamics is incompatible with conformal symmetry, an exceptional characteristic of Maxwell theory. As it happens, however, recent studies have found that there is a unique nonlinear extension of Maxwell's equations, dubbed \emph{ModMax}, which enjoys conformal symmetry and possesses duality rotation invariance at the same time~\cite{Bandos:2020jsw,Kosyakov:2020wxv}. It has also been shown to enjoy the discrete symmetry of Legendre self-duality~\cite{Danial-ModMax}. These properties motivate us to study radiative spacetimes sourced by ModMax matter.
The model is described by the following Lagrangian
\begin{equation}
	\mathbb{L}=-\cosh (\gamma)\,\mathcal{F}+\sinh (\gamma)\,\sqrt{\mathcal{F}^2+\mathcal{G}^2}.
\end{equation}

Finally, let us briefly mention the relevance of scalar matter within the context described above. On the one hand, Robinson-Trautman spacetimes sourced by scalar fields are much better behaved than their electromagnetic counterparts. For instance, they settle down to spherically symmetric solutions \cite{Tahamtan-PRD:2015,Tahamtan-PRD:2016}. On the other hand, scalar and electromagnetic fields are known to complement each other quite well in a variety of circumstances, for cases with nonlinear constitutive relations we refer the reader to Ref.~\cite{Cardenas:2014kaa,Tahamtan-PRD:2020, Tahamtan-Kundt:2017, Zhang:2021qga}. Which brings us to our final point, the RT geometries sourced by nonlinear sigma models of Ref.~\cite{Marcello-sigma}. A closely related, and somewhat unstudied, configuration with toroidal geometry which we analyze in order to go beyond spherical base manifolds in the context of stability analysis. Specifically, when performing the linear stability analysis we will use an expansion into Laplace eigenfunctions on a compact manifold (either torus or sphere) and evaluate whether the corresponding topology influences the results.

\section{Robinson--Trautman solution coupled to electromagnetism}




%
%

In this section, we study two NE models in RT class. The most general RT metric compatible with electromagnetic sources is given by
\begin{equation}\label{RT-metric}
\d s^2 = -(2H+Q)\,\d u^2-\,2\,\d u\,\d r + \frac{R^2}{{P}^2}\,(\d x^{2} + \d y^{2})
\end{equation} 
and we assume $u,r,x,y$ coordinate ordering. The non-shearing, non-twisting and expanding null geodesic congruence defining the RT geometry is given by $\partial_{r}$, $u$ is a retarded time with $u=const.$ being null hypersurfaces and the transversal space spanned by $x,y$ is customarily assumed to be of spherical topology (but planar versions with potential toroidal compactification are as well considered) with its geometry specified by $P(u,x,y)$. By introducing the $2H+Q$ metric function we keep the original form of the vacuum Robinson--Trautman spacetime with the metric function $2H = \Delta(\,\ln {P})-2r(\,\ln{P})_{,u} -{2m/r}$ explicitly included. Note that it is always possible to achieve $m=const.$ by coordinate transformation and function redefinitions in vacuum case. We assume $m$ to be zero, however one can always recover non-vanishing $m$ as a $1/r$ term in $Q$ appearing as an integration constant. The new metric function $Q=Q(u,r,x,y)$ is to be found based on the NE source. One can show that the above metric function $R=R(u,r)$ can be simplified into $R=r$ with a coordinate transformation \cite{Tahamtan-PRD:2021} and we assume such a situation from now on.

The nonzero electromagnetic field components compatible with the Robinson--Trautman class are $F_{ur}, F_{ux}, F_{uy}$ and $F_{xy}$. Since the metric \eqref{RT-metric} can accommodate only the outgoing rays aligned with the principle null direction $\partial_{r}$, we assume that $F_{xr}, F_{yr}$ or $F^{ux}, F^{uy}$ are zero since they would otherwise correspond to rays propagating in the opposite null direction (ingoing). This is related to fixing the initial conditions for the evolution of the Robinson--Trautman geometry which are usually assumed to be given on $u=u_{initial}$ hypersurface.

As shown in \cite{Tahamtan-PRD:2021} the Einstein tensor reduces to a single crucial component corresponding to the so-called Robinson--Trautman equation in the vacuum case (note that we have essentially hidden $m$ inside $Q$) 
 \begin{align}
 {G^{r}}_{u}=\frac{-1}{2r^2}\Delta\left(K+Q\right)-\frac{1}{r}\left[(\ln P)_{,u}\left(rQ_{,r}-2Q\right)+Q_{,u}\right]\,. \label{RT-equation} \nonumber\\
 \end{align}
Usually, all the other components of the Einstein equations are satisfied and the equation corresponding to \eqref{RT-equation} is crucial to determine whether solution is of type II (and therefore generic RT family member) or not (this was the case for several regular black hole models considered in \cite{Tahamtan-NE:2016, Tahamtan-PRD:2021}). Note that for trivial $Q=-\frac{\Lambda}{3}\,r^2$ (with $\Lambda$ being a cosmological constant) the Einstein equation corresponding to \eqref{RT-equation} with zero right-hand side reduces to $\Delta K=0$ which means that the solution is no longer type II but only type D.

\subsection{RegMax Lagrangian} \label{RegMax-magnetic}

In Ref.~\cite{Tahamtan-PRD:2021}, a new model of NE was found that is the only model which admits (both gravitationally and electromagnetically) radiative solutions for electrically charged black hole in the RT class. Independently, in Ref.~\cite{Tahamtan-PRD:2022}, this NE model was also discovered as a unique model of restricted NE (apart from the linear Maxwell theory) which provides slowly rotating black hole solutions with vector potential completely specified by the static solution. Later in \cite{Tahamtan2023}, we called this Lagrangian Regularized Maxwell (RegMax). The important feature of RegMax Lagrangian is the result we found in \cite{Tahamtan-PRD:2021}, where we showed that the electric radiative solution for RegMax Lagrangian are well posed around corresponding static solution, unlike for the analogical situation in the Maxwell case \cite{Kozameh_2008}. In the section IV--B of \cite{Tahamtan-PRD:2021} we have also mentioned that there are no magnetic radiative solutions in RT class for any theory other than Maxwell theory. However, this statement was based on a  mistake in the sign in equation (5.17) of \cite{Tahamtan-PRD:2021}. Although upon this correction one can not rule out the existence of radiative magnetic solutions in RT class there is still severe limitation for magnetically charged solutions. Namely, when NE models are considered for constructing magnetically sourced regular black holes another result from Ref.~\cite{Tahamtan-PRD:2021} applies, which rules out the presence of such radiative solutions in RT class. 

In order to explicitly show that there is a NE model admitting radiative magnetic solutions in RT class (albeit not providing a regular black hole), we use our RegMax Lagrangian corresponding to magnetic charge generalizing the static spherically symmetric case studied in \cite{Tahamtan2023}.

Thus, in order to find radiative magnetically charged solution in RT class for RegMax Lagrangian we have to generalize the Lagrangian to accommodate magnetic charge and ensure Maxwell limit in weak field regime as already discussed in \cite{Tahamtan2023}. Therefore, we redefine $s$, and flip the overall sign, to obtain the RegMax Lagrangian
\begin{equation}
	\mathbb{L}=8\,\alpha^4\,\left(1-3\ln(1-s)+\frac{s^3+3s^2-4s-2}{2(1-s)}\right)
\end{equation}
where $s=-{\left(\frac{\mathcal{F}}{2\,\alpha ^4}\right)^\frac{1}{4}}$ and we assume $\alpha$ to be positive. For details we refer to \cite{Tahamtan2023} (especially note that using absolute values a generally valid form can be given as discussed in the reference).

We assume the following field strength
\begin{eqnarray}
	\textbf{F}&=& \zeta(u,r,x,y)\, \d x \wedge \d y  \nonumber \\
	&&	+ \eta(u,r,x,y)\, \d u \wedge \d y + \xi(u,r,x,y)\, \d u \wedge \d x\,, \nonumber\\
\end{eqnarray}
which is consistent with RT geometry and corresponds to fields of a magnetic charge plus radiation. From \eqref{NE-MAX-TEN-2} one immediately concludes that $\zeta(u,r,x,y)$ is independent of $r$ (this is always valid no matter what is the model of NE). To simplify and compactify our notation we reparametrize one of the functions $\zeta=\frac{B}{P^2}$ introducing arbitrary function $B(u,x,y)$. From NE equations, i.e. \eqref{NE-MAX-TEN-2} and  \eqref{NE-MAX-TEN-1}, we see that both $\eta(u,r,x,y)$ and $\xi(u,r,x,y)$ should be $r$ independent as well. Moreover, we find
the following relations between the fields 
\begin{eqnarray}\label{NE-magnetic}
	\eta&=&\frac{\alpha\,B_{,x}}{2\,\sqrt{B}}\,,\quad\quad\quad \xi=-\frac{\alpha\,B_{,y}}{2\,\sqrt{B}}\ .
\end{eqnarray}
The independence of the radiative fields above is consistent with peeling behavior of radiation that should transfer energy to infinity. Indeed, if we calculate electromagnetic scalar $\phi_2$ we recover the correct decay in $r$
\[\phi_2=\frac{\alpha\,P}{\sqrt{2\,B}\,r}\left(B_{,y}+\mathrm{i} \,B_{,x}\right)\ ,\]
Finally, the following dynamical equation is coming from the modified Maxwell equation \eqref{NE-MAX-TEN-2}
\begin{equation}\label{Maxwell-constrain-magnetic charge}
	B\,{\Delta} \ln B+{\Delta} B=\frac{4\,B^{\frac{3}{2}}}{\alpha}\left((\ln B)_{,u}-2\,(\ln P)_{,u}\right)\ .
\end{equation}
Note that the electromagnetic field invariant for this case is
\[\mathcal{F}=\frac{2\,B^2}{r^4}\ .\]

We can find the form of the metric function $Q$ by using the "$rr$" component of the Einstein field equations \eqref{field equations}
\begin{eqnarray}
	Q(u,r,x,y)&=&\frac{22/3\alpha\,B(u,x,y)^{\frac{3}{2}}-2\,C(u,x,y)}{r}\nonumber\\
	&&-2\alpha^2 B(u,x,y)+4\alpha^3\,r \sqrt{B(u,x,y)}
 \nonumber\\
	&&-4\alpha^4 r^2\,\ln\left(1+\frac{\sqrt{B(u,x,y)}}{r\alpha}\right),\nonumber\\
\end{eqnarray}
where $C(u,x,y)$ is an integration constant. Similarly to \cite{Tahamtan-PRD:2021}, we can introduce $-2\,\mu(u,x,y)=22/3\alpha\,B(u,x,y)^{\frac{3}{2}}-2\,C(u,x,y)$, which gives the following asymptotic ($r\to\infty$) behavior
\begin{equation}
	Q \rightarrow -\frac{2\,\mu+4/3\,\alpha\,B^{\frac{3}{2}}}{r}+\frac{B^2}{r^2}+O\left(\frac{1}{r^3}\right)\ .
\end{equation}
From the "$rx$" and "$ry$" components of the field equations we arrive at the following restriction on coordinate dependence for the newly introduced function $\mu(u,x,y)=\mu(u)$. Subsequently, the dynamical equation $G^{r}{}_{u}=T^{r}{}_{u}$ simplifies into
\begin{eqnarray}
	\Delta{(K-2\alpha^2\,B)}+12\,\mu(\ln P)_{,u}-4\,\mu_{,u}=0 \ ,
\end{eqnarray}
having close resemblance to the vacuum RT equation and completely reproducing it for $B=0$.

 However we can make another, more natural, choice of the free functions. Since the RegMax model has Maxwell limit the solution should approach Reissner--N\"{o}rdstrom form asymptotically. In order to make this explicit 
 one can choose to reparametrize the integration constant $C(u,x,y)=m(u,x,y)+3\,\alpha\,B(u,x,y)^{3/2}$ to capture together all the terms behaving asymptotically as $\propto 1/r$ using newly introduced $m(u,x,y)$. In other words the metric function is then given by
\begin{eqnarray}
	Q(u,r,x,y)&=&-\frac{2\,m(u,x,y)-\frac{4}{3}\alpha\,B(u,x,y)^{3/2}}{r}\nonumber\\
 &&-2\alpha^2 B(u,x,y)+4\alpha^3\,r \sqrt{B(u,x,y)}\nonumber\\
 &&-4\alpha^4 r^2\,\ln\left(1+\frac{\sqrt{B(u,x,y)}}{r\alpha}\right),\nonumber
\end{eqnarray}
and the above solution has the following large $r$ expansion
\begin{equation}
Q \rightarrow -\frac{2\,m}{r}+\frac{B^2}{r^2}+O\left(\frac{1}{r^3}\right) \ .	
\end{equation}
Note that according to our assumptions the metric function $H$ is finite at the origin ($r=0$) and therefore the nature of the singularity (which resides at $r=0$) is determined by the sign of the divergent term in $Q$ which we denote $\tilde{M}=3\,m(u,x,y)-2\,\alpha\,B(u,x,y)^{3/2}$. The singularity is spacelike for $\tilde{M}>0$ and timelike for $\tilde{M}<0$. Similar situation arises also for static spherically symmetric case discussed in \cite{Tahamtan2023} where $3\,m-2\alpha\,B^{\frac{3}{2}}>0$ had to be satisfied in order to ensure spacelike singularity (so-called Schwarzschild-type behavior). As we will see, the nature of the singularity has crucial significance for the stability and well-posedness.


From "$rx$" and "$ry$" components of the field equations we get elliptic type constraint
\begin{equation}\label{m-MadMax-mass}
	\Delta{m}=\frac{\alpha\,\sqrt{B}}{2}\,\left(3\,\Delta{B}-B\,\Delta{(\ln B)}\right) \ .
\end{equation}
And finally the modified RT equation becomes
\begin{eqnarray}\label{RT-MAdMax-Magnetic}
	\Delta{K}+12\,m\,(\ln P)_{,u}-4\,m_{,u}=\frac{\alpha^2\,P^2}{B}\left(B^2_{,x}+B^2_{,y}\right)\ .
\end{eqnarray}
Thus the specific solution is obtained by solving dynamical equations \eqref{Maxwell-constrain-magnetic charge}, \eqref{RT-MAdMax-Magnetic} and satisfying constraint \eqref{m-MadMax-mass}.

\subsection{ModMax}


Let us now turn our attention to the ModMax model \cite{Bandos:2020jsw, Kosyakov:2020wxv, Danial-ModMax}
\begin{equation}
	\mathbb{L}=-\cosh (\gamma)\,\mathcal{F}+\sinh (\gamma)\,\sqrt{\mathcal{F}^2+\mathcal{G}^2} \ .
\end{equation}
Since it shares important properties with the Maxwell theory (duality and conformal invariance) it is interesting to see if the drawbacks of the Maxwell theory in the RT class (ill-posedness) are cured by nonlinearity in ModMax theory. In order to fully exploit the appearance of the pseudoinvariant $\mathcal{G}$ in the Lagrangian we consider both electric and magnetic charge. The most general electromagnetic field compatible with RT class in such a case is given by
 \begin{eqnarray}
 	\textbf{F}&=&-E(u,r,x,y)\, \d u \wedge \d r + \zeta(u,r,x,y)\, \d x \wedge \d y  \nonumber \\
 &&	+ \eta(u,r,x,y)\, \d u \wedge \d y + \xi(u,r,x,y)\, \d u \wedge \d x \nonumber\\ \ ,
 \end{eqnarray}
where by using \eqref{NE-MAX-TEN-2} we conclude that $\zeta(u,r,x,y)$ is independent of $r$ and we again consider $\zeta=\frac{B}{P^2}$ with general $B(u,x,y)$. By using one component of \eqref{NE-MAX-TEN-1} we find the form of $E$ 
\begin{equation}\label{Electric Field-ModMax}
	E(u,r,x,y)=\frac{e^{-\gamma}\,A_e(u,x,y)}{r^2} \ .
\end{equation}
From the remaining modified Maxwell equations, \eqref{NE-MAX-TEN-2} and  \eqref{NE-MAX-TEN-1} we uncover the form of radiation fields $\xi,\eta$ and "electric" and "magnetic" fields 
\begin{eqnarray}
&&\eta(u,r,x,y)=\epsilon_{0}(u,x,y)\,,\quad \xi(u,r,x,y)=\epsilon_{1}(u,x,y) \nonumber
\\
&&A_{e}(u,x,y)=q_e(u)\,,\quad B(u,x,y)=q_m(u) \ .
 \end{eqnarray}
As in the previous NE model, the radiation fields are independent of $r$ leading to correct $r$ dependence in $\phi_{2}$ consistent with transfer of energy to infinity by electromagnetic radiation. The close similarity of ModMax model to the Maxwell theory is explicitly visible on the form of electromagnetic field invariants  
\[\mathcal{F}=\frac{2\,\left(q_m^2-e^{-2\gamma}\,q_e^2\right)}{r^4}\ ,\quad \mathcal{G}=\frac{4\,e^{-\gamma}\,q_{e}\,q_{m}}{r^4}\ ,\]
with the only difference being the exponential factor.

From the "$rr$" components of the field equations \eqref{field equations} we find the metric solution "$Q$"
 \begin{equation}\label{ModMax-metric}
 	Q(u,r,x,y)=-\frac{2\,m(u,x,y)}{r}+\frac{\left(q_e(u)^2+q_m(u)^2\right)\,e^{-\gamma}}{r^2}\ ,
 \end{equation}
where we denoted the constant of integration as $-2\,m(u,x,y)$ to yield the proper Schwarzschild limit. Substituting the form of the metric given by \eqref{ModMax-metric} into $G^{r}{}_{x}-T^{r}{}_{x}=0$ we get the following result 
\begin{equation}
	m_{,x}=2\,\Gamma\,\left(q_m\,\epsilon_{0}-\,q_e\,e^{-\gamma}\,\epsilon_{1}\right)\ ,
\end{equation}
where
\[\Gamma=\frac{\left(q_m^2+q_e^2\right)}{q_m^2\,e^{\gamma}+q_e^2\,e^{-\gamma}}\ .\]
Similarly, from  $G^{r}{}_{y}-T^{r}{}_{y}=0$ we obtain
\begin{equation}
	m_{,y}=-2\,\Gamma\,\left(q_m\,\epsilon_{1}+\,q_e\,e^{-\gamma}\,\epsilon_{0}\right)\ .
\end{equation}
Combining these equations we arrive at
\[\Delta m=-2\,\Gamma \,P^2\, \left[q_m\,(\epsilon_{1,y}-\epsilon_{0,x})-\,q_e\,e^{-\gamma}\,(\epsilon_{1,x}+\epsilon_{0,y})\right]\ .\]
Obviously this equation is expressed in terms of the radiative fields. It is possible to use the two remaining components of the modified Maxwell equations to obtain dynamical eqyation for $\Delta m$
\begin{equation}\label{ModMax-m}
\Delta m=-4\,e^{-\gamma}\,(\ln P)_{,u} \,\left(q_m^2+q_e^2\right)+e^{-\gamma}\,\left(q_m^2+q_e^2\right)_{,u}\ .
\end{equation}

And finally, the last equation  $G^{r}{}_{u}=T^{r}{}_{u}$ can be split into two equations for different orders in $r$. One is equivalent to \eqref{ModMax-m} and the other has the following form
\begin{align}
	\Delta K+12m\,(\ln P)_{,u}-4m_{,u}=&4\,\Gamma\,P^2\left\{(\epsilon_{0})^2+(\epsilon_{1})^2\right\},\label{RT-ModMax-eq}\ .
\end{align}

All the final expressions and equations for the ModMax model are close to Maxwell theory with the only differences given by the appearance of factor $e^{\gamma}$ as expected from the experience of spherically symmetric solutions. Especially notice that due to the form of metric function $Q$ \eqref{ModMax-metric} the singularity is always timelike, similarly to the Maxwell case. That is why already at this stage one suspects that the results regarding well-posedness will be unchanged compared to the Maxwell case.

\section{Robinson--Trautman solution with sigma Model}
The nonlinear sigma model coupled to RT geometry provided in \cite{Marcello-sigma} is described by the following action
\begin{equation}
    S_{\sigma}=\frac{\kappa}{4}\int \d x^4 \sqrt{-g}\, \mathrm{Tr}(R^{\mu}R_{\mu}) 
\end{equation}
where $\kappa>0$ is a coupling constant and $R_{\mu}$ is the Maurer-Cartan 1-form given by 
\begin{equation}
    R_{\mu}=U^{-1}\,\nabla_{\mu}U
\end{equation}
in terms of a $SU(2)$ valued scalar field specifying the degrees of freedom of the nonlinear sigma model.

Corresponding energy momentum tensor can be given in the form
\begin{equation}\label{EMT-sigma}
    T_{\mu\nu}=\frac{\kappa}{2}\mathrm{Tr}\left(R_{\mu}R_{\nu}-\frac{1}{2}g_{\mu\nu}R^{\alpha}R_{\alpha}\right)\ ,
\end{equation}
while the equtions of motion for the sigma field are
\begin{equation}
    \nabla_{\mu}R^{\mu}=0\ .
\end{equation}

Adopting the standard parametrization using Pauli matrices we can write
\begin{equation}
    R_{\mu}=\mathrm{i}\,R_{\mu}^{j}\,\tau_{j}
\end{equation}
where 
\begin{equation}
    R_{\mu}^{i}=\epsilon^{ijk}U_{j}\nabla_{\mu}U_{k}+U^{0}\nabla_{\mu}U^{i}-U^{i}\nabla_{\mu}U^{0}
\end{equation}
and $U_{i}$ are parametrizing our $SU(2)$ valued scalar field and therefore have to satisfy $(U^{0})^{2}+U_{i}U^{i}=1$.

The specific sigma model solution in RT class considered in \cite{Marcello-sigma} uses the standard RT metric nsatz
\begin{equation}
	\d s^2 = -2H\,\d u^2-\,2\,\d u\,\d r + \frac{r^2}{{P}^2}\,(\d x^{2} + \d y^{2}),
\end{equation}
where $H(u,r,x,y)$ and $P(u,x,y)$. The potential corresponding to sigma model field was assumed in the following form 
\begin{eqnarray}
	U_0=\frac{\cos x}{\sqrt{2}}, \,\qquad\qquad	U_1=\frac{\sin y}{\sqrt{2}}\nonumber\\
	U_2=\frac{\cos y}{\sqrt{2}}, \,\qquad\qquad 	U_3=\frac{\sin x}{\sqrt{2}}
\end{eqnarray}
which corresponds to a toroidal mapping of the transversal coordinates $x,y$.
The explicit form of the energy momentum tensor \eqref{EMT-sigma} then takes the following simple form 
\begin{equation}\label{SigmaSource}
T^{\mu}{}_{\nu}=diagonal\left[-\frac{\kappa\,P^2}{2\,r^2}, -\frac{\kappa\,P^2}{2\,r^2},0,0\right]\ .
\end{equation}

The metric solution then takes the form  
 \begin{equation}\label{Sigma-H}
	2H=\Delta(\,\ln {P})-\frac{\kappa\,P^2}{2}-2r(\,\ln{P})_{,u}-\frac{2\,m}{r},
\end{equation}
where $\Delta=P^2\,(\partial_{xx}+\partial_{yy})$. The only remaining nontrivial Einstein equation is the (modified) RT equation
\begin{align}\label{RT-SigmaModel}
	\Delta (K-\frac{\kappa\,P^2}{2})+12m\,(\ln P)_{,u}-4m_{,u}=0,
\end{align}
where $K=\Delta(\,\ln {P})$ and $\Delta m=0$. Note that although the sigma model metric solution is no longer vacuum it still admits arbitrary rescaling of the retarded time coordinate as described in \cite{Stephanietal:book} (but not the additional transformation in transversal coordinates). And similarly to vacuum case, it can be used to set $m(u)$ to constant without limiting the generality of the solution. Due to the form of metric function $H$ \eqref{Sigma-H} the singularity is always spacelike.

The above solution is in general a dynamical one in RT class. It is interesting to see how this spacetime behaves in the static case, analogically to vacuum RT which has Schwarzschild black hole as the static case. If we assume that all metric functions appearing in \eqref{RT-SigmaModel} are constant in $u$ we obtain condition   
\begin{equation}\label{Laplace-Sigma}
    \Delta \left(K-\frac{\kappa\,P^2}{2}\right)=0
\end{equation}
Assuming that the two-dimensional geometry of $u=const.,r=const.$ subspaces is compact and remembering that harmonic functions on compact manifolds are constants we see that in general the expression in bracket in \eqref{Laplace-Sigma} should be a constant which leads to the following equation for $P$
\begin{equation}\label{Sigma-static}
(\partial_{xx}+\partial_{yy})\,(\ln{P})=\kappa/2+k/P^2\ .
\end{equation}
When $P=1$ and $k=-\kappa/2$ we obtain static solution discussed in \cite{Marcello-sigma} as a toroidal AdS black hole (upon trivial inclusion of cosmological constant in the derivation above). 

For $k=0$ we obtain another nontrivial static solution. Since it will not be discussed in the following study of well-posedness we are providing its description in appendix \ref{appendix:sigma-solution}.

\section{Linear perturbations around stationary solution and well-posedness}

In this section, we study linearized versions of the dynamical equations governing the evolution of solutions discussed in the previous sections. We determine stability and well-posedness of the linear system and comment on possible extensions to non-linear regime using PDE theory, as in Ref.~\cite{Gustafsson-Kreiss-Oliger}.
 
In this paper, we found RT solutions corresponding to three different matter sources. First, we obtained a magnetically charged solution considering the RegMax NE Lagrangian. Although the metric solution is similar to the electric case (whose stability and well-posedness was analyzed in Ref.~\cite{Tahamtan-PRD:2021}), the electromagnetic fields (especially the radiative part) are quiet different. Therefore we want to ensure that previous results indeed generalize to the magnetic case.

The second solution we obtained was for the ModMax NE model in Eq. \eqref{ModMax-metric}. In this case, we consider both both electric and magnetic charge. This configuration is a radiative generalization of the black hole in Ref.~\cite{Danial-ModMax}. However, due to its close similarity to the Maxwell-RT solution we expect the same unfavorable results, e.g., similar to those obtained in Ref.~\cite{Kozameh_2008}. 

Finally, in order to investigate RT spacetimes beyond spherical symmetry we have considered a nonlinear sigma model sources. Our solution concerning a sigma model with toroidal RT geometry is reported in the appendix. In this section, we analyse instead the solution found in Ref.~\cite{Marcello-sigma}, which we have discussed above. In order to carry out the procedure we modify it to account for the change in topology.

The method we use for studying the linear stability and well-posedness follows the approach in \cite{Kozameh_2008}. We assume perturbation around corresponding static spherically symmetric solution which we evolve using the RT-matter coupled equations. 

The function $P$ determining the geometry of transversal two-spaces is considered in the following form,
\begin{equation}\label{pertub-P}
	P=P_{0}(1+g)\ ,
\end{equation}
in which $P_0$ corresponds to symmetric situation of a static solution and $g(u,x,y)$ represents a perturbation. Therefore we linearize the field equations by neglecting terms with higher orders of $g$ and its spatial derivatives.

In the case of RegMax and ModMax NE models the function $P_0=1+1/4(x^2+y^2)$ corresponds to spherical symmetry and we assume that perturbations can be expanded into spherical harmonics. Subsequently we work with individual modes labeled by $l$, thereby obtaining 
\begin{equation}\label{spherical harmonic-g}
    \tilde{\Delta}\,g=-l(l+1)g
\end{equation}
where $\tilde{\Delta}\equiv P^{2}_{0}(\partial_{xx}+\partial_{yy})$. 
Based on \eqref{pertub-P}, the following is valid up to linear order
\begin{equation}\label{pertub-K}
    K=\Delta \ln P \simeq 1+2g+\tilde{\Delta}g
\end{equation}
where the absolute term represents Gaussian curvature of a sphere.

For sigma model the static geometry is no longer spherically symmetric but toroidal instead. We therefore have $P_0=1$ and assume that perturbations can be expanded into eigenfunctions of a Laplace operator on a flat torus, thus obtaining 
\begin{equation}\label{torus-harmonic}
    \tilde{\Delta}\,g=-\lambda^2\,g\ ,
\end{equation}
where the eigenvalue satisfies $\lambda^{2}=\lambda_{x}^{2}+\lambda_{y}^{2}$ for positive integers $\lambda_{x,y}$ corresponding to fundamental frequencies in given directions that are determined by radii of the torus. We neglect zero eigenvalue since it corresponds to constant solution which can be absorbed into a redefinition of $P_{0}$.

\subsection{Stability of RegMax}\label{subsection:RegMax}

Based on the discussion of magnetically charged RegMax solution in \ref{RegMax-magnetic} we have two time-evolution equations --- one from the modified Maxwell equation, \eqref{Maxwell-constrain-magnetic charge}, and one from the RT equation \eqref{RT-MAdMax-Magnetic}. Additionally, there is a constraint on $m$ from RT equation, namely \eqref{m-MadMax-mass}. Similarly to linearization applied to the function $P$ determining transversal geometry, \eqref{pertub-P}, we assume linear perturbation using spherical harmonics around static values for the magnetic field $B$ and for $m$ 
\[B=B_0(1+\bar{B}),\qquad m=m_{0}+\psi\ ,\]
where
\[\tilde{\Delta}\,\bar{B}=-l(l+1)\,\bar{B},\quad \tilde{\Delta}\,\psi=-l(l+1)\,\psi\ .\]
If we apply the above described linearization to \eqref{m-MadMax-mass} we obtain the following
\begin{equation}
\tilde{\Delta} \psi=\alpha\,B^{\frac{3}{2}}_0\,\tilde{\Delta} \bar{B}	
\end{equation}
and by taking time derivative and further simplifications we obtain relation for retarded time derivatives 
\begin{equation}
	\psi_{,u}=\alpha\,B^{\frac{3}{2}}_0\,\bar{B}_{,u}\ .
\end{equation}
Using the above relation together with \eqref{pertub-P} and \eqref{pertub-K}, after inserting \eqref{spherical harmonic-g} into \eqref{Maxwell-constrain-magnetic charge} and  \eqref{RT-MAdMax-Magnetic} we find the following two equations
\begin{eqnarray}
	\bar{B}_{,u}&=&\sigma\, l(l+1)\,\left[j\,\bar{B}+\Omega\,g\right]	\\\nonumber\\
	g_{,u}&=&\sigma\, l(l+1)\, \left[\alpha^2\,B_0\,\bar{B}+\frac{\Omega}{2}\,g \,\right]	
\end{eqnarray}
that determine the (retarded) time evolution of the linearized system. The matrix form of this system is
\begin{equation}\label{RegMax-matrix}
\begin{bmatrix} 
	\bar{B}_{,u}\\\\
	g_{,u}
\end{bmatrix} 
=\sigma\, l(l+1)\,
\begin{bmatrix} 
	j && \Omega  \\\\
	\alpha^2\,B_0 && \frac{\Omega}{2} \\
\end{bmatrix}
\begin{bmatrix} 
	\bar{B}\\\\
	g
\end{bmatrix} 
\end{equation}

where $\Omega=l(l+1)-2$, $j=\frac{3\alpha\,m_0}{\sqrt{B_0}}$ and $\sigma=-\frac{1}{2\,(3\,m_0-2\alpha\,B^{\frac{3}{2}}_0)}$. Note that the following condition holds
\begin{equation}\label{constants-rel}
3\,m_0-2\alpha\,B^{\frac{3}{2}}_0>0\ ,
\end{equation}
for the values of the static solution (around which the analysis is performed) if we consider so called Schwarzschild-type solution where the singularity is spacelike. While for Reissner--Nordstr\"{o}m-type static solutions the inequality is reversed and singularity is timelike.

Naturally, we assume exponential solution ($\sim \e^{\nu u}$) given by the eigenvalues of the matrix on the right-hand side of \eqref{RegMax-matrix}. These eigenvalues are 
\begin{equation}\label{RegMax-eigenvalues}
	\nu_{\pm}=\frac{\sigma\, l(l+1)}{4}\left(\Omega+2\,j \pm \sqrt{\left(\Omega-2\,j\right)^2+16\,\Omega\,\alpha^2\,B_0}\right).
\end{equation}
Both real eigenvalues $\nu_{\pm}$ are negative for all $l\geq 2$ (lower modes can be removed by coordinate transformations) thanks to \eqref{constants-rel}. First, this means that both branches of the solution for perturbation are decaying exponentially fast resulting in linear stability of the solution. And second, the sign of the eigenvalues (especially for $l\gg 1$) together with correct dependence on mode number means that the linear problem is well-posed. Moreover, the eigenvalues have a nonzero gap away from zero which can be computed by putting $l=2$ into $\nu_{-}$ for fixed parameters $\alpha, m_0, B_0$ satisfying \eqref{constants-rel}. Using the results in \cite{Gustafsson-Kreiss-Oliger} one can extend the linear well-posedness towards local well-posedness around the static solution for the nonlinear problem. Namely, one can work in "frozen coefficients formulation" and utilize the fact that the principal symbol of the PDE system is strongly parabolic for any background solution satisfying \eqref{constants-rel}.

Inverting the inequality in \eqref{constants-rel} and thus considering background solution with timelike singularity (Reissner--Nordstr\"{o}m type) one clearly obtains eigenvalues of opposite sign and therefore the problem becomes unstable and ill-posed. This demonstrates that analysis of parabolic problems can in general only be local in the vicinity of some solution. 

Additionally note, that since the important factor $3\,m_0-2\alpha\,B^{\frac{3}{2}}_0$ \eqref{constants-rel} appears in the denominator of eigenvalues \eqref{RegMax-eigenvalues}, very small positive and negative values of this factor lead to eigenvalues that are large. This means that solutions very close to the opposite sides of the threshold between timelike and spacelike singularity cases produce behaviors that are arbitrarily far from each other (if measured by the eigenvalues).

\subsection{Stability of ModMax}

We now move on to the ModMax NE model, in this case, the two time evolution equations are derived from \eqref{ModMax-m} and \eqref{RT-ModMax-eq}. We assume $q_m=q_e=q_0$ for simplicity in \eqref{ModMax-m} which leads to
\begin{equation}\label{evol-1}
	\left(\ln P\right)_{,u}=-\frac{\Delta m\,e^{\gamma}}{8\,q_{0}^2}\, .  
\end{equation}
By substituting the above expression into \eqref{RT-ModMax-eq} we arrive at the second time evolution equation, 
\begin{eqnarray}\label{evol-2}
&	m_{,u}=\frac{\Delta K}{4}-\frac{3\,m\,e^{\gamma}}{8\,q_{0}^2}\,\Delta m-\Gamma\,P^2\left\{(\epsilon_{0})^2+(\epsilon_{1})^2\right\}.\nonumber\\
\end{eqnarray}

Following the procedure from the previous section we assume linear perturbation for $m$,
\[m=m_{0}+\psi\ ,\ \tilde{\Delta}\,\psi=-l(l+1)\,\psi\ .\]
Neglecting all higher-order terms in $g$, $\psi$ and their derivatives, the equations \ref{evol-1} and \ref{evol-2} lead to the following linearized system
\begin{eqnarray}\label{evol-1-2}
	\psi_{,u}&=&\frac{3\,l(l+1)\,m_{0}\,e^{\gamma}}{8\,q_{0}^2}\,\psi+\frac{l(l+1)\,\Omega}{4}\,g\ ,\nonumber \\
	g_{,u}&=&\frac{l(l+1)\,e^{\gamma}}{8\,q_{0}^2}\psi \ .
\end{eqnarray}
Such system provides exponential solution for both  $g$ and $\psi$ ($\sim \e^{\nu u}$) determined by eigenvalues
\begin{equation}
	\nu_{\pm}=\frac{l(l+1)\,e^{\gamma}}{16\,q_{0}^2}\left(3\,m_0 \pm \sqrt{9m_0^2\,e^{\gamma}+8\,q_{0}^2\,\Omega} \right)\ .
\end{equation}
Clearly, the $\nu_{+}$ eigenvalue stays always positive while $\nu_{-}$ is negative from certain value of $l$ depending on the constants of the solution. This means the system is linearly unstable (in both directions of retarded time evolution) and the whole RT--ModMax system is as well clearly not well-posed since $\lim_{l\to\infty}\nu_{+}=\infty$. This is in complete agreement with the results for RT--Maxwell system obtained in \cite{Kozameh_2008} which is not surprising considering the minor modification which the ModMax NE model brings into this picture (e.g., singularity is timelike for both).

\subsection{Stability of Sigma Model}
We consider $m(u)$ as a constant, $m_0$, based on the remarks after equation \eqref{RT-SigmaModel}. We again assume the function $P$ decomposed as in \eqref{pertub-P} but now with $P_0=1$ because of toroidal geometry and we use \eqref{torus-harmonic}.

Using \eqref{pertub-P}, the following linearization holds
\[K=\Delta \ln P \simeq 1+2g+\tilde{\Delta}g\]
where $\tilde{\Delta}\equiv (\partial_{xx}+\partial_{yy})$. 

The only dynamical equation in this case is \eqref{RT-SigmaModel} and after applying the linearization the only evolution equation reads
\[g_{,u}=-\frac{\lambda^{2}\left[\lambda^{2}+\kappa-2\right]}{12\,m_0}\,g\]
and has a single eigenvalue
\begin{equation}
    \nu = -\frac{\lambda^{2}\left[\lambda^{2}+\kappa-2\right]}{12\,m_0}\ .
\end{equation}
This eigenvalue is negative for all considered $\lambda$ and therefore the solution decays exponentially in $u$. It is also linearly well-posed since $\lim_{\lambda \to\infty}\nu=-\infty$. And all eigenvalues satisfy $\nu\geq -\frac{\kappa}{6m_{0}}$ which assures that the linear well-posedness can be extended to local nonlinear one around corresponding static solution (using the techniques of \cite{Gustafsson-Kreiss-Oliger} mentioned at the end of \ref{subsection:RegMax}). Let us repeat here that the singularity is always spacelike for this solution.

Unlike the previous two cases we did not discuss the equation governing the source. The reason for this is an apparent "rigidity" of the source which stays the same no matter what is the specific RT geometry determined by $P$.

\section{Conclusion}

In this paper, we have continued the study of radiative Robinson--Trautman spacetimes sourced by nonlinear matter models. For an important class of models in nonlinear electrodynamics, it was previously established that there is only one nonlinear extension of Maxwell's equations which produces a well-posed problem. Electrically charged radiative solutions of this type were the first to be studied in this Regularized Maxwell model. However, since that theory lacks electric-magnetic duality invariance then magnetic solutions cannot be trivially obtained from electric ones. In this work, we fill this gap and provide the first magnetically sourced radiative spacetimes. The stability and well-posedness for these solutions crucially depends on the character of the singularity of the background solution (as in the case of previously studied electrically charged solutions) which can be both spacelike or timelike depending on the parameters. In order to search for new potentially well-posed systems we consider the Modified Maxwell nonlinear theory which possesses duality invariance and conformal symmetry. Theories such as these escape many established results. Motivated by this we have constructed the first Robinson--Trautman configurations with both electric and magnetic charge for that matter model. However, their resemblance with the Maxwell case (namely the timelike character of singularity) leads to ill-posed initial value problems. We have based our analysis on linear perturbation theory around static solutions and properties of the principal symbols for the PDE systems under consideration. Our exploration includes the examination of radiative systems sourced by gauged nonlinear sigma models. For these spacetimes we have extended the standard analysis and so are able to consider non-spherical topology. These solutions are stable and well-posed while their singularity is spacelike.

Based on the obtained results we argue for the validity of the conjecture \ref{conjecture}. Namely, that the physically relevant black hole solutions (those that are stable and well-posed) in the systems of RT geometry coupled with matter sources are those that contain spacelike singularity as opposed to timelike one. This seems consistent with expected behavior close to Cauchy horizon present in the cases with timelike singularities, namely perturbative instability. Of course, one should strive to prove the conjecture rigorously which will be the topic of our future research.

Although in the presence of matter fields we do not have global existence results like those available  for the vacuum case \cite{Chrusciel:1991vxx,Chrusciel:1992tj}, we can reasonably assume that generic solutions in the well-posed cases would settle down to static solution as $u\to\infty$ and thus $u=\infty$ would correspond to event horizon of the static solution. That means that although we analyzed the solution only in the exterior part of the spacetime (above outer event horizon) the results are sensitive to the nature of the singularity that is hidden beyond the horizon. Evidently, more extensive analysis of the global solutions is needed in order to make definite conclusions.

\begin{acknowledgments}
T.T. and O.S.  are grateful to D. Hilditch for many helpful discussions during visit to Centra (IST, Lisbon) and online. It is our informal contribution to theoretical investigations related to gravitational waves within the LISA Consortium. D.F-A. is supported by Agencia Nacional de Investigaci\'{o}n y Desarrollo under FONDECYT grant No. 3220083. T.T. is grateful for support from  GA\v{C}R 23-07457S  and  O.S.  for support from GA\v{C}R 22-14791S grant of the Czech Science Foundation. 
\end{acknowledgments}

\appendix
\section{Electrically charged RegMax Solution}

Here, we provide the electrical analogue of the magnetic solution pertinent to this work. The matter Lagrangian is
\begin{equation}
	\mathbb{L}=-8\,\alpha^4\,\left(1-3\ln(1-s)+\frac{s^3+3s^2-4s-2}{2(1-s)}\right)
\end{equation}
where $s=\left(\frac{-\mathcal{F}}{2\,\alpha^4}\right)^\frac{1}{4}$, whereas, the field is given by
\begin{eqnarray}
 	\textbf{F}&=&-E(u,r,x,y)\, \d u \wedge \d r  \nonumber \\
 &&	+ \eta(u,r,x,y)\, \d u \wedge \d y + \xi(u,r,x,y)\, \d u \wedge \d x \nonumber\\ \ ,
 \end{eqnarray}
with
\begin{equation}
	E=\frac{\alpha^2\,A}{(r\,\alpha+\sqrt{A})^2}
\end{equation}
\begin{eqnarray}
	\eta&=&-\frac{r^2 \alpha^3\,A_{,y}}{2\,\sqrt{A}\,(r\,\alpha+\sqrt{A})^2}\,,
	\\
	 \xi&=&-\frac{r^2 \alpha^3\,A_{,x}}{2\,\sqrt{A}\,(r\,\alpha+\sqrt{A})^2}\,. 
\end{eqnarray}
Notice that calculating the electromagnetic scalar
\[\phi_2=\frac{r\,\alpha^3\,P}{\sqrt{2\,A}\,(r\,\alpha+\sqrt{A})^2}\left(A_{,x}-\mathrm{i} \,A_{,y}\right)\]
reveals it to behave asymptotically as $1/r$. Also, observe that the last equation from NE equations is, in this case,
\begin{equation}
		A\,{\Delta} \ln A+{\Delta} A=\frac{4\,A^{\frac{3}{2}}}{\alpha}\left((\ln A)_{,u}-2\,(\ln P)_{,u}\right).
\end{equation}
From the Einstein equations we have that
\begin{eqnarray}
	&&	Q(u,r,x,y)=-\frac{2\,\mu(u)}{r}-2\alpha^2 A(u,x,y)\nonumber\\
	&&+4\alpha^3\,r \sqrt{A(u,x,y)}-4\alpha^4 r^2\,\ln\left(1+\frac{\sqrt{A(u,x,y)}}{r\alpha}\right),\nonumber\\
\end{eqnarray}
and
\begin{eqnarray}
	\Delta{(K-2\alpha^2\,A)}+12\,\mu(\ln P)_{,u}-4\,\mu_{,u}=0.
\end{eqnarray}


In order to have the mass from in RN limit we have to redefine the mass term 
\begin{eqnarray}
	&&	Q(u,r,x,y)=-\frac{2\,\tilde{m}(u,x,y)}{r}-2\alpha^2 A(u,x,y)\nonumber\\
	&&+4\alpha^3\,r \sqrt{A(u,x,y)}-4\alpha^4 r^2\,\ln\left(1+\frac{\sqrt{A(u,x,y)}}{r\alpha}\right),\nonumber\\
\end{eqnarray}
where $\tilde{m}=m(u,x,y)-2/3\,\alpha\,A(u,x,y)^{\frac{3}{2}}$ and
\[\Delta{m}=-\frac{\alpha\,\sqrt{A}}{2}\,\left(A\,\Delta{(\ln B)}-3\,\Delta{A}\right)\]
yielding the desired result
\begin{eqnarray}
	\Delta{K}+12\,m\,(\ln P)_{,u}-4\,m_{,u}=\frac{\alpha^2\,P^2}{A}\left(A^2_{,x}+A^2_{,y}\right).
\end{eqnarray}

\section{New static sigma model solution}\label{appendix:sigma-solution}
Here, we discuss a new solution sourced by a sigma model and with RT geometry. For $k=0$ we obtain another nontrivial static solution of equation \eqref{Sigma-static} given by 
\begin{multline}\label{P-drop}
    P(x,y)=\\
    \exp{\left(\frac{\kappa}{8}((1+e)\,x^2+(1-e)\,y^2)+C_1 x+C_2y+C_{0}\right)}
\end{multline}
where the integration constants $C_{0,1,2}$ can be absorbed into trivial redefinitions of the coordinates $x,y$ and from now on we can assume they have these values: $C_{0}=0, C_{1}=0, C_{2}=0$. On the other hand the constant $e$ cannot be absorbed without changing relative scaling of coordinates $x,y$ and thus has a true geometrical meaning. For $e=0$ we have circularly symmetric $P$ while for nonzero values we obtain elliptic deformations.

\begin{figure}[t]
  \centering
\includegraphics[scale=0.6]{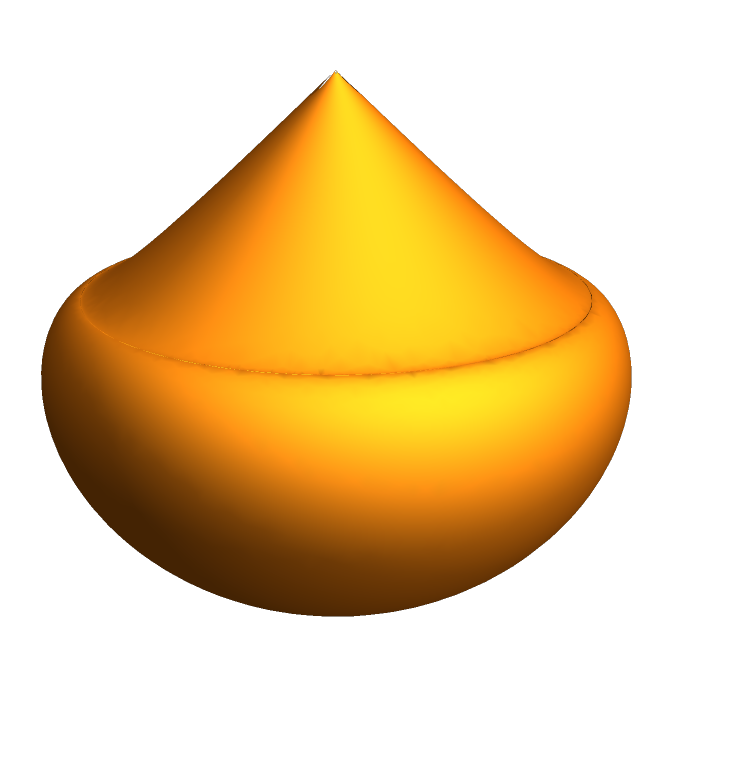}
 \caption{The embedding of $P_{drop}$ for $\kappa=1$ with the seam connecting plotting routines for $\frac{\kappa}{8}(\,x^2+\,y^2) \gtrless 1$.}  \label{drop}
\end{figure}

For simplicity let us analyze the $e=0$ case of \eqref{P-drop}. Close to the origin ($x,y \ll 1$) we have the following approximation
\begin{equation}
    P_{drop}=\exp{\left(\frac{\kappa}{8}(\,x^2+\,y^2)\right)} \sim 1+\frac{\kappa}{8}(\,x^2+\,y^2) + O(x^4,y^4)
\end{equation}
which has the form of metric on a sphere in stereographic coordinates. Further away from the origin we will have deviations and we can visualize the complete geometry using embedding into three-dimensional Euclidean space as shown in figure \ref{drop}.

Further let us note that the geometry given by solution \eqref{P-drop} is the following
\begin{equation}\label{metric-drop}
	\d s^2 = \frac{2\,m}{r}\,\d u^2-\,2\,\d u\,\d r + \frac{r^2}{{P_{drop}}^2}\,(\d x^{2} + \d y^{2}).
\end{equation}
This can again be made into an AdS black hole (by generalizing with cosmological constant) with a weird horizon given by embedding \ref{drop} and having a singularity on the vertical half-axis apart from the Schwarzschild-like one at the origin as evident from the Kretschmann scalar
\begin{equation}
    {\frac {{{\rm e}^{\frac{\kappa}{2} \left( {x}^{2}+{y}^{2} \right) }}{\kappa}^{2}{r}^{2}+8\,{{\rm e}^{\frac{\kappa}{4} \left( {x}^{2}+{y}^{2} \right) }}\kappa\,m\,r+48\,{m}^{2}\\
\mbox{}}{{r}^{6}}}
\end{equation}
The singular half-axis is evidently caused by the sigma model field and disappears if $\kappa=0$.

Alternatively, one may observe that signs of different terms in the metric correspond to values inside the Schwarzschild black hole where the spacetime is of cosmological model type (Kantowski--Sachs). Therefore one expects this to be the case for \eqref{metric-drop} as well. Indeed, by performing the following coordinate transformation
\begin{equation}
   u=\left(\frac{3}{4}\right)^{\frac{4}{3}}m^{-\frac{1}{3}}\tau^{\frac{4}{3}}+\rho, r=\left(\frac{9}{2}\right)^{\frac{1}{3}}m^{\frac{1}{3}}\tau^{\frac{2}{3}}, x=\tilde{x}, y=\tilde{y}
\end{equation}
we arrive at
\begin{align}\label{metric-Kasner}
	\d s^2 =& -\d\tau^{2} + \left(\frac{4}{3}\right)^{\frac{2}{3}}m^{\frac{2}{3}}\tau^{-\frac{2}{3}}\,\d \rho^2 \\
 &+ \left(\frac{9}{2}\right)^{\frac{2}{3}}m^{\frac{2}{3}}\tau^{\frac{4}{3}}\frac{1}{{P_{drop}(\tilde{x},\tilde{y})}^2}\,(\d \tilde{x}^{2} + \d \tilde{y}^{2})\ .
\end{align}
Interestingly, the anisotropic expansion factors of the form $\tau^{2\, p_{i}}$ satisfy the Kasner conditions $\sum p_{i}=1$ and $\sum p_{i}^{2}=1$. Therefore the spacetime behaves like a deformed Kasner solution with initial cosmological singularity at $\tau=0$ and additional singularity on each $\tau=const.>0$ hypersurface caused by the divergence in $P_{drop}$ which appears due to the sigma model field.

\bibliography{References}

\end{document}